\documentclass[preprint,12pt]{elsarticle}
\journal{Physics Letter B}

\begin{document}
\begin{frontmatter}
\title{The $\rho-\omega$ splitting in constituent quark models}

\author{J. Vijande}
\address{Departamento de F\' \i sica At\'omica, Molecular y Nuclear e IFIC,
Universidad de Valencia - CSIC, E-46100 Burjassot, Valencia, Spain}
\author{A. Valcarce}
\address{Departamento de F\'\i sica Fundamental, 
Universidad de Salamanca, E-37008 Salamanca, Spain}

\begin{abstract}
In this letter we present a solution to describe simultaneously the light
isoscalar and isovector vector mesons in constituent quark models.
In Ref.~\cite{Vij05} the $q\bar q$ spectrum was studied in a 
generalized constituent quark model
constrained by the $NN$ phenomenology and the baryon spectrum. An
overall good fit to the available experimental data was obtained. A 
major problem of this description was the relative position of the vector
$\omega$ and $\rho$ mesons. The present results
improve the description of the isoscalar meson spectroscopy.
They should serve as a step forward in distinguishing
conventional quark model mesons from exotic states.
\end{abstract}

\end{frontmatter}
Meson spectroscopy is probably one of the best laboratories
to learn about the properties of QCD. Its simplicity,
a two-body problem, has been masked by
the recent advent of a large number of states that 
may fit into the description of mesons made of more
than two particles, multiquarks~\cite{Jaf07}. Such a 
complication was guessed long ago on the first adventures in the
intriguing sector of the light scalar mesons. It seems
nowadays unavoidable to resort to higher Fock space
components to tame the bewildering landscape drawn
by experiment. To understand the nature
of new resonances it is important that we have a 
template to compare observed states 
with theoretical predictions. To this respect,
phenomenological models are an important 
theoretical tool to study meson properties. 

Ref.~\cite{Vij05} derived a quark-quark interaction to study
the complete meson spectra, from the light-
to the heavy-quark sector. The idea behind that calculation was to 
contribute to the identification of
possible exotic, non quark-antiquark, structures. For this purpose one needs
a pattern to compare with. Then, one can go through the different puzzling states
to decide whether a discrepancy is most likely a problem 
with the theoretical model or whether it may
signal some new physics. 
Although obtaining an overall good agreement between theory and experiment, 
a major problem was noticed on the description of the isovector
and isoscalar vector mesons. This difficulty was specially evident
for the ground states, fitting the $\rho (770)$ to have a 
mass of 772 MeV, we got a mass of 691 MeV for the $\omega (782)$.

This result seemed to reborn an already observed deficiency of constituent quark
models, the lack of an isospin-dependent mechanism in the 
light quark sector. This was first observed when studying the nucleon-nucleon 
($NN$) scattering: while the isosinglet phase shifts were described in a very 
accurate manner, the isotriplet ones could not be correctly 
reproduced~\cite{Val05}. This was solved through 
an isospin-dependent mechanism, the coupling to $N\Delta$ channels,
giving a nice description of the $^1S_0$ partial wave~\cite{Val95}
and also of the $^3P_J$ ones~\cite{Dai97}. Other calculations
have remedied this problem by considering an isospin-dependent scalar
exchange~\cite{Mac87}. When dealing with  
the meson spectroscopy the landscape offered by constituent
quark models is even worst. On the one hand, the one-gluon exchange 
does not distinguish between isoscalar and isovector
mesons~\cite{Bha81}. On the other hand, pseudoscalar
boson exchanges generate the reverse ordering 
as compared to experiment~\cite{Vij05}.
Although meson-loop contributions would definitely help to remedy
such a discrepancy, the $\omega-\rho$ mass difference
induced by the non-strange pseudoscalar exchanges seems to be too
large ($\approx 100$ MeV) to be explained by such a mechanism~\cite{Pic99}. 
We will return to this discussion later on.
In this letter we present a mechanism missed in our original
work that provides with the correct isospin dependence.

Let us briefly resume the basic lines of the quark model used.
Although heavy mesons are properly described by nonrelativistic potential 
models reflecting the dynamics expected from QCD:
a universal one-gluon exchange plus a linear confining 
potential~\cite{Eic75}, the dynamics of the light-quark sector 
is expected to be dominated by the nonperturbative 
spontaneous breaking of chiral symmetry.
Chiral symmetry suggests to divide the quarks into two 
different sectors: light quarks ($u$, $d$, and $s$) where
chiral symmetry is spontaneously broken, and heavy quarks ($c$ and $b$) 
where the symmetry is explicitly broken. The origin of the constituent quark 
mass can be traced back to the spontaneous breaking 
of chiral symmetry and consequently constituent quarks
interact through the exchange of Goldstone bosons~\cite{Man84},
these exchanges being essential to obtain a correct description of the $NN$
phenomenology and the light baryon spectrum~\cite{Val05}. 
Beyond the chiral symmetry breaking scale one expects the dynamics
being governed by QCD perturbative effects, taken into account
through the one-gluon-exchange (OGE) potential.
Therefore, for the light quark sector hadrons can be described as
systems of confined constituent quarks (antiquarks) interacting through
gluon and boson exchanges, whereas for the heavy quark sector hadrons 
are systems of confined current quarks interacting through gluon exchanges. 
Finally, any model imitating QCD should incorporate 
confinement, that takes into account that the
only observed hadrons are color singlets. The only indication
we have about the nature of confinement is through lattice QCD studies,
showing that $Q\bar Q$ systems are well reproduced at short 
distances by a linear potential. Such a potential can be physically
interpreted in a picture where the quark and the antiquark are linked
with a one-dimensional color flux tube. The spontaneous creation of
light-quark pairs may give rise to a breakup of the color flux tube.
It has been proposed that this translates into a screened 
potential~\cite{Bal01}, in such a way that the potential does not
rise continuously but it saturates at some interquark distance.

Once perturbative (one-gluon exchange) and nonperturbative (confinement
and chiral symmetry breaking) aspects of QCD are considered, one
ends up with a quark-quark interaction of the form
(we refer to a light quark,
$u$ or $d$, as $n$, $s$ is used for the strange quark and $Q$ for the
heavy quarks $c$ and $b$):
\begin{equation}
V_{q_iq_j}=\left\{ 
\begin{array}{ll}
(q_iq_j)=(nn,ns,ss) \Rightarrow V_{CON}+V_{OGE}+V_{\chi} &  \\ 
(q_iq_j)=(nQ,QQ) \Rightarrow V_{CON}+V_{OGE} &  
\end{array}
\right.  \label{pot}
\end{equation}
where $V_{\chi}$ stands for the chiral part of the interaction we 
will discuss below. See Ref. \cite{Vij05}
for a thorough discussion of the model and explicit expressions of
the potentials.

Let us now center on the chiral part of the interaction.
The constituent quark mass appears because of the spontaneous
breaking of the original $SU(3)_{L}\otimes SU(3)_{R}$ chiral symmetry at
some momentum scale. The interaction of the quarks with the
medium modifies the propagator of light quarks
and they acquire a momentum dependent mass which drops to zero
for large momenta. The momentum dependent quark mass acts 
as a natural cutoff of the
theory. In this domain of momenta, a simple Lagrangian
invariant under the chiral transformation can be derived as~\cite{Dia86}

\begin{equation}
L=\overline{\psi}({\rm i} \gamma^\mu \partial_\mu -m_q\, U^{\gamma _{5}})\psi
\label{equ1}
\end{equation}
being $U^{\gamma _{5}}=\exp (i\pi ^{a}\lambda ^{a}\gamma_{5}/f_{\pi })$. 
$\pi ^{a}$ denotes the pseudoscalar fields $(\vec{\pi }
,K_{i},\eta_8)$ with i=1,...,4, and $m_q$ is the quark mass. In the presence
of instanton fluctuations of QCD, quarks acquire a dynamical mass which 
depends upon the momentum transfer $Q$ as illustrated in Ref.~\cite{Dia86}
(see Fig. 4 of this reference). This momentum dependence shows a good 
correspondence with lattice data~\cite{Bow04} (see Fig. 6 of this reference)
and it can be parametrized as $m(Q^2)=m(0)F(Q^2)$ with
\begin{equation}
F(Q^2)=\left[ \frac{\Lambda^{2}}{\Lambda^{2}+Q^{2}} \right] ^{\frac{1}{2}}
\label{eq2}
\end{equation}
where $\Lambda$ determines the scale at which chiral symmetry is
broken. Once a constituent quark mass is generated such particles have
to interact through Goldstone modes. Whereas the Lagrangian
$\overline{\psi} (i\gamma^\mu \partial_\mu -m_q)\psi$ is not invariant under
chiral rotations, that of Eq. (\ref{equ1})
is invariant since the rotation
of the quark fields can be compensated renaming the bosons fields.
$U^{\gamma _{5}}$ can be expanded in terms of boson fields as,    
\begin{equation}
U^{\gamma _{5}}=1+\frac{{\rm i}}{f_{\pi }}\gamma ^{5}\lambda ^{a}\pi ^{a}-
\frac{1}{2f_{\pi }^{2}}\pi ^{a}\pi ^{a}+...
\label{eq3}
\end{equation}
The first term generates the constituent quark mass and the second one gives
rise to a one-boson exchange interaction between quarks. The main
contribution of the third term comes from the two-pion exchange which 
was simulated in Ref.~\cite{Vij05} by means of single scalar exchange potential. 
Therefore, the chiral part of the quark-quark interaction in Ref. ~\cite{Vij05}
can be resumed as follows,
\begin{equation}
V_{qq}(\vec r_{ij}) \, = \,
V_{\pi}(\vec r_{ij}) \, + \,
V_{\sigma}(\vec r_{ij}) \, + \,
V_{K}(\vec r_{ij}) \, + \,
V_{\eta}(\vec r_{ij}) \, ,
\label{tt}
\end{equation}

With this model we presented in Ref. \cite{Vij05} results for all 
mesons. Let us note that in Eq. (\ref{tt}) we made use of a simplification 
that we were not aware could be playing a significant role for the meson
spectroscopy. As already mentioned we used
a single isoscalar-scalar exchange between quarks instead of the full octet of 
scalar mesons. When we applied this model to the study of the baryon-baryon 
interaction~\cite{Gar07}, we noticed for the first time the 
deficiency we face on Ref.~\cite{Vij05}: the need of an isospin-dependent 
scalar interaction. This was the reason why
we analyzed in Ref.~\cite{Gar07} the effect of including the
full scalar octet. These scalar potentials have
the same radial form and a different $SU(3)$ operatorial dependence, i.e.,

\begin{equation}
V ({\vec r}_{ij}) =
V_{a_0} ({\vec r}_{ij}) \sum_{F=1}^3 \lambda_i^F \cdot \lambda_j^F +
V_\kappa ({\vec r}_{ij}) \sum_{F=4}^7 \lambda_i^F \cdot \lambda_j^F +
V_{f_0} ({\vec r}_{ij}) \lambda_i^8 \cdot \lambda_j^8 +
V_{\sigma} ({\vec r}_{ij}) \lambda_i^0 \cdot \lambda_j^0
\label{eq1}
\end{equation}

\noindent
where
\begin{equation}
V_{k} ({\vec r}_{ij}) = - {g^2_{ch} \over {4 \, \pi}} \,
{\Lambda_{k}^2 \over \Lambda_{k}^2 - m_{k}^2}
\, m_{k} \, \left[ Y (m_{k} \, r_{ij})-
{\Lambda_{k} \over {m_{k}}} \,
Y (\Lambda_{k} \, r_{ij}) \right] \, ,
\label{OSE}
\end{equation}

\noindent
with $k=a_0, \kappa, f_0 $ or $\sigma$. The model parameters $g^2_{ch}/4\pi=0.54$, $m_q=$ 313 MeV,
$\Lambda_{\sigma}=4.2$ fm$^{-1}$, $m_{\sigma}=3.42$ fm$^{-1}$, ... have been taken from Ref.~\cite{Vij05}.
$g_{ch}=m_q/f_{\pi}$ is the quark-meson coupling constant and $Y(x)$ the standard Yukawa function defined
by $Y(x)=e^{-x}/x$.
Such interaction could also
help in solving the problems observed in the description of the
meson spectra. Therefore we have redone the study of the
meson spectra considering the full scalar octet as indicated
in Eq.~(\ref{eq1}). In order
to avoid the uncertainties associated to the masses and cut-offs
of the different scalar mesons we will make use of the most economical
way, keeping a single radial structure and preserving the different
operatorial dependence,

\begin{equation}
\label{gen-sigma}
V ({\vec r}_{ij}) =
V_{\sigma} ({\vec r}_{ij}) \left( \beta \sum_{F=1}^3 \lambda_i^F \cdot \lambda_j^F +
\delta \sum_{F=4}^7 \lambda_i^F \cdot \lambda_j^F +
\gamma \lambda_i^8 \cdot \lambda_j^8 +
\alpha \lambda_i^0 \cdot \lambda_j^0 \right)
\end{equation}
where $V_{\sigma}$ is a the one-sigma exchange potential of Ref.~\cite{Vij05}.

Among the possible criteria that can be used to fit the new parameters
appearing in Eq. (\ref{gen-sigma}) we have chosen to leave unaffected 
the isovector meson sector, minimizing in this way the changes 
respect to the original parametrization~\cite{Vij05}. 
Once this is done, the parameters are related through 
\begin{equation}
\label{rela}
\beta+{{\gamma}\over3}+{\alpha{2\over3}}=1 \, .
\end{equation}

By fitting the $\omega-\rho$ mass splitting we obtained $\beta=0.75$. 
Thus, any value of $\alpha$ and $\gamma$ fulfilling
$\gamma+2\alpha=3/4$ gives rise to the correct $\omega-\rho$ mass splitting. 
These two parameters have been determined through a global fit of the strange meson sector, 
obtaining $\alpha=0.47$ and $\gamma=-0.19$. A value of $\delta=-0.60$ has been fitted to 
the $\eta-\eta'$ mass splitting. Only two parameters of the original parametrization
obtained in Ref.\cite{Vij05} have been modified, namely the
value of the quark-gluon strong coupling constant, $\alpha_s$, in the light-strange 
($\alpha_{us}=0.475$) and strange-strange ($\alpha_{ss}=0.523$) sectors.
Let us note that the values used for the QCD strong coupling constant are
in good agreement with its expected momentum dependence~\cite{Shi97}
when the typical momentum scale for each flavor sector is assimilated
to the reduced mass of the interacting pair~\cite{Hal93}.

We show in Table~\ref{t1} the results obtained for those states whose
masses are modified when considering the full scalar octet exchange
compared to those of Refs.~\cite{Kol00,Sil97,God85,Ono82,Bha81}.
Our election of the different scalar octet strengths
does not modify the masses of the isovector mesons. 
The flavor dependence of the scalar octet does reverse the order
of the $\rho$ and $\omega$ meson masses in a perfect agreement 
with experiment. No other important modifications are observed,
only small variations of masses but without reversing any other
set of states. This mechanism is only active in the light and strange
sector, and therefore heavy meson masses are also not modified.
It is also important to notice that the $\omega$ meson arises from
ideal mixing with a nonzero hidden strangeness component, 
and therefore the model describing its interaction should incorporate $SU(3)$
effects, in other words the flavor content of the full scalar octet.

\begin{table}[tbp]
\caption{Meson masses in MeV. Experimental data (Exp.) are taken
from Ref.~\protect\cite{Yao06}. The two numbers given in the fourth column correspond to
two different models.}
\label{t1}
\begin{center}
\begin{tabular}{|cccc|ccccc|}
\hline
State & Ref.~\cite{Vij05} & This work & Exp. & Ref.~\cite{Kol00} & Ref.~\cite{Sil97} & Ref.~\cite{God85}
& Ref.~\cite{Ono82} & Ref.~\cite{Bha81} \\ 
\hline
$\pi$              & 139 & 139 & 138 & 138/140 & 141  &  150 & 132 & 136  \\ 
$\rho$             & 772 & 772 & 776 & 778/785 & 771  &  770 & 774 & 777 \\ 
$\omega$           & 691 & 783 & 783 & 778/785 & 771  &  780 & 774 & 777 \\ 
$K$                & 496 & 495 & 495 & 499/506 & 499  &  470 & 501 & 520 \\ 
$K^*$              & 910 & 912 & 893 & 870/890 & 891  &  900 & 900 & 905 \\ 
$\phi$             &1020 & 1018& 1019& 954/990 & 965  & 1020 &1011 &1017  \\
$\eta$             & 572 & 547 & 548 & 531/533 & 141  &  520 & 132 & 136  \\
$\eta'$            & 956 & 958 & 958 & 975/950 & $-$  &  960 & 696 & $-$ \\
\hline
\end{tabular}
\end{center}
\end{table}

Although our phenomenological model is certainly effective in solving
remaining discrepancies on the meson spectroscopy, there is still room
for the contribution of other mechanisms that have demonstrated to be
very important for the $\rho-\omega$ mass splitting and their
strong decays. They are the meson-loop contributions mentioned
above that have been extensively discussed in 
the literature, see Ref.~\cite{Pic99} for a general overview. 
Contributions arising from mixing of hadrons with multiple-hadron 
states were proposed as a solution to the problem
in models where $\rho$ and $\omega$ would be degenerate due 
to the absence of $SU(3)$ flavor breaking interactions.
The $\rho^0$ and the $\omega$
have a similar quark structure differing in their total
isospin, and hence $G-$parity. As mixing between the $\omega-$meson
and two-pion states is forbidden by $G-$parity, the mass of the $\omega-$meson
does not receive such a contribution. Ref.~\cite{Pic99} found a mass splitting
of the order of 25 MeV from the contribution of 
$\pi\pi$, $K\overline K$, $\omega\pi$, $\rho\pi$, $\omega\eta$, $\rho\eta$
and $KK^*$ intermediate channels. The conclusion of this work
is that the inclusion of two-meson loops change the charge radius
of the $\rho$ meson by only about 10 per cent, but they are very
important for the $\rho-\omega$ mass splitting and their strong decays.
Thus such a mixing with multiple-hadron states, that has been mocked up
in our phenomenological approach in the fitted parameters, would clearly
help to improve the performance of the model for specific processes
or phenomena. 

In summary, we have found an isospin dependent mechanism that allows
to solve the remaining discrepancies, beyond the scalar mesons, in
the description of meson spectroscopy in constituent quark models.
The relevance of this mechanism was first observed in the study of 
the baryon-baryon interaction, giving coherence to the study of
the low energy hadron spectra and their interactions in the framework
of constituent quarks models. 

\section*{Acknowledgments}
This work has been partially funded by Ministerio de Ciencia y Tecnolog\'{\i}a
under Contract No. FPA2007-65748 and by EU FEDER, and by Junta de Castilla y Le\'{o}n
under Contracts No. SA016A17 and GR12.

\end{document}